\documentclass[11pt]{socc14}

\usepackage{amsmath}
\usepackage{mathptmx}

\usepackage{graphicx}
\usepackage{caption}
\usepackage{subcaption}
\usepackage{balance}  
\usepackage{pdfpages}
\usepackage{fancyhdr}
\usepackage{exscale}
\usepackage{booktabs}
\usepackage{multicol}
\usepackage{multirow}
\usepackage{amsfonts}
\usepackage{amssymb}
\usepackage{ntheorem}
\usepackage{paralist}
\usepackage{verbatim}
\usepackage{color}
\usepackage{hyperref}
\usepackage{enumitem}
\usepackage{listings}      
\usepackage{algorithm}
\usepackage{algorithmicx}  
\usepackage[noend]{algpseudocode} 
\usepackage{setspace}
\usepackage{tikz}
\usepackage{array}
\usepackage{relsize}
\usepackage{diagbox}
\usepackage{listings}

\definecolor{darkred}{rgb}{0.5,0,0}
\definecolor{darkgreen}{rgb}{0,0.5,0}
\definecolor{darkblue}{rgb}{0,0,0.5}

\hypersetup{
    pdfstartview={FitV},
    pdftitle={},
    pdfauthor={},       
    pdfsubject={},        
    pdfcreator={\TeX maker},  
    pdfproducer={\LaTeX},   
    pdfkeywords={},
    pdfnewwindow=true,
    colorlinks=true,
    linkcolor=darkred,
    citecolor=darkgreen,
    urlcolor=darkblue
}

\definecolor{Red}{rgb}{1.0,0,0}
\definecolor{Blue}{rgb}{0,0,1.0} 
\definecolor{Green}{rgb}{0,1.0,0}

\lstdefinelanguage{scala}{
  morekeywords={abstract,case,catch,class,def,%
    do,else,extends,false,final,finally,%
    for,if,implicit,import,match,mixin,%
    new,null,object,override,package,%
    private,protected,requires,return,sealed,%
    super,this,throw,trait,true,try,%
    type,val,var,while,with,yield},
  otherkeywords={=>,<-,<\%,<:,>:,\#,@},
  sensitive=true,
  morecomment=[l]{//},
  morecomment=[n]{/*}{*/},
  morestring=[b]",
  morestring=[b]',
  morestring=[b]"""
}

\lstdefinelanguage{stream}{
  morekeywords={filter, map, flatMap, source, sink,
              join, fold, groupBy,trigger, cogroup, evict, trigger, cross, project, union, scan, window, eviction},
  otherkeywords={=>,<-,<\%,<:,>:,\#,@},
  sensitive=true,
  morecomment=[l]{//},
  morecomment=[n]{/*}{*/},
  morestring=[b]",
  morestring=[b]',
  morestring=[b]"""
}
 
\algblockdefx[Trigger]{Trigger}{EndTrigger}
    [2][event]{\textbf{trigger} $\langle#1$ $\mid$ $#2\rangle$;}
    
\algnewcommand\algorithmicfe{\textbf{ for each }}
\algnewcommand\algorithmicas{\textbf{ as }}

\algblockdefx[ForEach]{ForEach}{EndForEach}[2][collection]
   {\algorithmicfe $#1$ \textbf{as} $#2$}

\algblockdefx[UponEvent]{Upon}{EndUpon}
  [2][event]{\textbf{upon event} $\langle#1$ $\mid$ \emph{#2}$\rangle$ \textbf{do} }

\usetikzlibrary{shapes.geometric}
\usetikzlibrary{arrows}

\makeatletter
\def\@copyrightspace{\relax}
\makeatother
\begin{document}

%





\authorinfo{Paris Carbone$^1$~~~~~~~~
Gyula F\'{o}ra$^2$~~~~~~~~
Stephan Ewen$^3$~~~~~~~~
Seif Haridi$^{1,2}$~~~~~~~~Kostas Tzoumas$^3\vspace{20pt}$\\ 
  {$^1$KTH Royal Institute of Technology - \{parisc,haridi\}@kth.se}\\
  {$^2$Swedish Institute of Computer Science  - \{gyfora, seif\}@sics.se} \\
 {$^3$Data Artisans GmbH   - \{stephan, kostas\}@data-artisans.com}
}

\title{Lightweight Asynchronous Snapshots for Distributed Dataflows}

%

\maketitle

\begin{abstract}
Distributed stateful stream processing enables the deployment and execution of large scale continuous computations in the cloud, targeting both low latency and high throughput. One of the most fundamental challenges of this paradigm is providing processing guarantees under potential failures. Existing approaches rely on periodic global state snapshots that can be used for failure recovery. Those approaches suffer from two main drawbacks. First, they often stall the overall computation which impacts ingestion. Second, they eagerly persist all records in transit along with the operation states which results in larger snapshots than required. In this work we propose Asynchronous Barrier Snapshotting (ABS), a lightweight algorithm suited for modern dataflow execution engines that minimises space requirements. ABS persists only operator states on acyclic execution topologies while keeping a minimal record log on cyclic dataflows. We implemented ABS on Apache Flink, a distributed analytics engine that supports stateful stream processing. Our evaluation shows that our algorithm does not have a heavy impact on the execution, maintaining linear scalability and performing well with frequent snapshots.

\end{abstract}


\section{Introduction}
\label{intro}
Distributed dataflow processing is an emerging paradigm for data intensive computing which allows continuous computations on data in high volumes, targeting low end-to-end latency while guaranteeing high throughput. Several time-critical applications could benefit from dataflow processing systems such as Apache Flink \cite{apacheflink} and Naiad \cite{murray2013naiad}, especially in the domains of real-time analysis (e.g. predictive analytics and complex event processing). Fault tolerance is of paramount importance in such systems, as failures cannot be afforded in most real-world use cases. Currently known approaches that guarantee exactly-once semantics on stateful processing systems rely on global, consistent snapshots of the execution state. However, there are two main drawbacks that make their application inefficient for real-time stream processing. Synchronous snapshotting techniques stop the overall execution of a distributed computation in order to obtain a consistent view of the overall state. Furthermore, to our knowledge all of the existing algorithms for distributed snapshots include records that are in transit in channels or unprocessed messages throughout the execution graph as part of the snapshotted state. Most often this includes state that is larger than required.

In this work, we focus on providing lightweight snapshotting, specifically targeted at distributed stateful dataflow systems, with low impact on performance. Our solution provides asynchronous state snapshots with low space costs that contain only operator states in acyclic execution topologies. Additionally, we cover the case of cyclic execution graphs by applying downstream backup on selected parts of the topology while keeping the snapshot state to minimum. Our technique does not halt the streaming operation and it only introduces a small runtime overhead. The contributions of this paper can be summarised as follows:

\begin{itemize}
		\item We propose and implement an asynchronous snapshotting algorithm that achieves minimal snapshots on acyclic execution graphs.
	\item We describe and implement a generalisation of our algorithm that works on cyclic execution graphs.
	\item We show the benefits of our approach compared to the state-of-the-art using Apache Flink Streaming as a base system for comparisons.
\end{itemize}

\noindent The rest of the paper is organised as follows: Section \ref{bg} gives an overview of existing approaches for distributed global snapshots in stateful dataflow systems. Section \ref{flink} provides an overview of the Apache Flink processing and execution model followed by Section \ref{abs} where we describe our main approach to global snapshotting in detail. Our recovery scheme is described briefly in Section \ref{recov}. Finally, Section \ref{impl} summarises our implementation followed by our evaluation in Section \ref{eval} and future work and conclusion in Section \ref{futurework}.

\section{Related Work}
\label{bg}

Several recovery mechanisms have been proposed during the last decade for systems that do continuous processing \cite{chandy1985distributed, murray2013naiad}. Systems that emulate continuous processing into stateless distributed batch computations such as Discretized Streams and Comet \cite{zaharia2012discretized,he2010comet} rely on state recomputation. On the other hand, stateful dataflow systems such as Naiad, SDGs, Piccolo  and SEEP \cite{murray2013naiad,fernandez2014making,power2010piccolo,castro2013integrating} , which are our main focus in this work, use checkpointing to obtain consistent snapshots of the global execution for failure recovery.
 
The problem of consistent global snapshots in distributed environments, as introduced by Chandy and Lamport \cite{chandy1985distributed}, has been researched extensively throughout the last decades \cite{chandy1985distributed,lai1987distributed,kshemkalyani1995introduction}. A global snapshot theoretically reflects the overall state of an execution, or a possible state at a specific instance of its operation. A simple but costly approach employed by Naiad \cite{murray2013naiad} is to perform a synchronous snapshot in three steps: first halting the overall computation of the execution graph, then performing the snapshot and finally instructing each task to continue its operation once the global snapshot is complete. This approach has a high impact on both throughput and space requirements due to the need to block the whole computation, while also relying on upstream backup that logs emitted records at the producer side.  Another popular approach, originally proposed by Chandy and Lamport \cite{chandy1985distributed}, that is deployed in many systems today is to perform snapshots asynchronously while eagerly doing upstream backup \cite{chandy1985distributed,low2012distributed,fernandez2014making}. This is achieved by distributing markers throughout the execution graph that trigger the persistence of operator and channel state. This approach though still suffers from additional space requirements due to the need of an upstream backup and as a result higher recovery times caused by the reprocessing of backup records.  Our approach extends the original asynchronous snapshotting idea of Chandy and Lamport, however, it considers no backup logging of records for acyclic graphs while also keeping very selective backup records on cyclic execution graphs.

\section{Background: The Apache Flink System}
\label{flink}
Our current work is guided by the need for fault tolerance on Apache Flink Streaming, a distributed stream analytics system that is part of the Apache Flink Stack (former Stratosphere \cite{alexandrov2014stratosphere}). Apache Flink is architectured around a generic runtime engine uniformly processing both batch and streaming jobs composed of stateful interconnected tasks. Analytics jobs in Flink are compiled into directed graphs of tasks. Data elements are fetched from external sources and routed through the task graph in a pipelined fashion. Tasks are continuously manipulating their internal state based on the received inputs and are generating new outputs.

\subsection{The Streaming Programming Model}
\label{model}
The Apache Flink API for stream processing allows the composition of complex streaming analytics jobs by exposing unbounded partitioned data streams (partially ordered sequences of records) as its core data abstraction, called \emph{DataStreams}. \emph{DataStreams} can be created from external sources (e.g. message queues, socket streams, custom generators) or by invoking operations on other \emph{DataStreams}. \emph{DataStreams} support several operators such as \emph{map}, \emph{filter} and \emph{reduce} in the form of higher order functions that are applied incrementally per record and generate new \emph{DataStreams}. Every operator can be parallelised by placing parallel instances to run on different partitions of the respective stream, thus, allowing the distributed execution of stream transformations.

The code example in \ref{wc_code} shows how to implement a simple incremental word count in Apache Flink. In this program words are read from a text file and the current count for each word is printed to the standard output. This is a stateful streaming program as sources need to be aware of their current file offset and counters need to maintain the current count for each word as their internal state.

\begin{figure}[t!]
\centering
\includegraphics[width=1.8in]{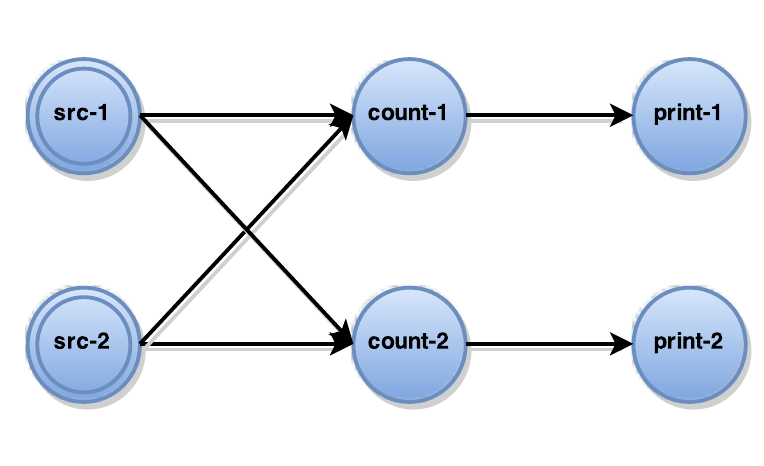}
\vspace{-1em}
\caption{The execution graph for incremental word count}
\label{fig_execgraph}
\end{figure}

\lstset{language=Scala, }  
\renewcommand{\lstlistingname}{Example}      
\begin{lstlisting}[frame=single,caption=Incremental Word Count, label=wc_code, captionpos = b]  
val env : StreamExecutionEnvironment = ...
env.setParallelism(2)

val wordStream = env.readTextFile(path)
val countStream = wordStream.groupBy(_).count
countStream.print
\end{lstlisting}


\begin{figure*}[!t]
\centering
\includegraphics[width=7in]{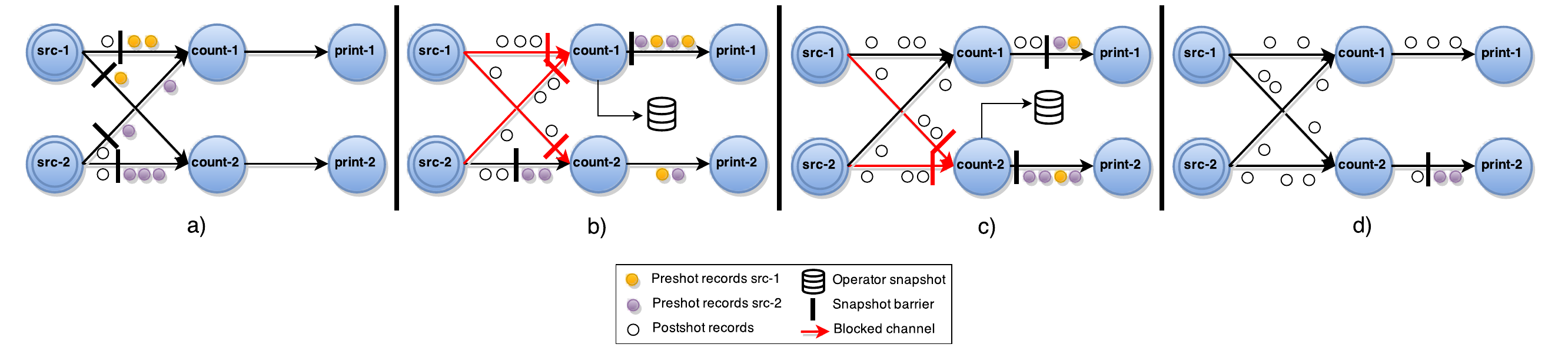}
\vspace{-1em}
\caption{Asynchronous barrier snapshots for acyclic graphs}
\label{abs-dag}
\end{figure*}

\subsection{Distributed Dataflow Execution}
\label{execution}

When a user executes an application all \emph{DataStream} operators compile into an execution graph that is in principle a directed graph $G = (T,E)$, similarly to Naiad \cite{murray2013naiad}  where vertices $T$ represent tasks and edges $E$ represent data channels between tasks. An  execution graph is depicted in Fig. \ref{fig_execgraph} for the incremental word count example. As shown, every instance of an operator is encapsulated on a respective task. Tasks can be further classified as sources when they have no input channels and sinks when no output channels are set. Furthermore, $M$ denotes the set of all records transferred by tasks during their parallel execution. Each task $t \in T$ encapsulates the independent execution of an operator instance and is composed of the following: (1) a set of input and output channels: $I_t, O_t \subseteq E$; (2) an operator state $s_t$ and (3) a user defined function (UDF) $f_t$. Data ingestion is pull-based : during its execution each task consumes input records, updates its operator state and generates new records according to its user defined function. More specifically, for each record $r \in M$ received by a task $t \in T$ a new state $s'_t$ is produced along with a set of output records $D \subseteq M$ according to its UDF $f_t : s_t, r \mapsto \langle s'_t, D \rangle$.

\section{Asynchronous Barrier Snapshotting}
\label{abs}

In order to provide consistent results, distributed processing systems need to be resilient to task failures. A way of providing this resilience is to periodically capture snapshots of the execution graph which can be used later to recover from failures. A snapshot is a global state of the execution graph, capturing all necessary information to restart the computation from that specific execution state. 
\subsection{Problem Definition}

We define a global snapshot $G^* = (T^*, E^*)$ of an execution graph $G = (T,E)$ as a set of all task and edge states, $T^*$ and $E^*$ respectively. In more detail, $T^*$ consists of all operator states $s^*_t \in T^*, \forall t \in T$, while $E^*$ is a set of all channel states $e^* \in E^*$ where $e^*$ consists of records that are in transit on $e$. 

We require that certain properties hold for each snapshot $G^*$ in order to guarantee correct results after recovery such as \emph{termination} and \emph{feasibility} as described by Tel \cite{tel2000introduction}. 

\noindent \emph{Termination} guarantees that a snapshot algorithm eventually finishes in finite time after its initiation if all processes are alive.  \emph{Feasibility} expresses the meaningfulness of a snapshot, i.e. that during the snapshotting process no information has been lost regarding the computation. Formally, this implies that causal order \cite{lamport1978time} is maintained in the snapshot such that records delivered in tasks are also sent from the viewpoint of a snapshot.

\begin{algorithm}[h]
\caption{Asynchronous Barrier Snapshotting for Acyclic Execution Graphs} 
\label{carbonefora}
\begin{algorithmic}[1]
\Upon[Init]{input\_channels,  output\_channels, fun, init\_state}
  \State $state :=  init\_state;$ $blocked\_inputs := \emptyset;$
  \State $inputs := input\_channels;$
  \State $outputs := output\_channels;$ \emph{udf} $:= fun;$
\EndUpon
\Upon[receive]{input, $\langle barrier \rangle$}
  \If{$input \neq Nil$}
  \State $blocked\_inputs := blocked\_inputs \cup \{input\};$
  \State \textbf{trigger} $\langle$block $\mid$ input$\rangle$;\label{lst:line:block}
  \EndIf
  \If{$blocked\_inputs = inputs$} 
    \State $blocked\_inputs := \emptyset;$
    \State \textbf{broadcast} $\langle$send $\mid$ outputs, $\langle barrier \rangle$$\rangle$;\label{lst:line:barrier}
    \State \textbf{trigger} $\langle$snapshot $\mid$ state$\rangle$;\label{lst:line:snap}
     \ForEach[inputs]{input}
      \State \textbf{trigger} $\langle$unblock $\mid$ input $\rangle$;\label{lst:line:unblock}
    \EndForEach    

  \EndIf
 \EndUpon
\Upon[receive]{input, msg}
    \State $\{state', out\_records\} :=$ \emph{udf}$(msg, state);$
    \State $state := state';$
    \ForEach[out\_records]{\{output, out\_record\}}
      \State \textbf{trigger} $\langle$send $\mid$ output, out\_record$\rangle$;
    \EndForEach
\EndUpon
\end{algorithmic}
\end{algorithm}

\subsection{ABS for Acyclic Dataflows}
\label{cabs}

It is feasible to do snapshots without persisting channel states when the execution is divided into stages. Stages divide the injected data streams and all associated computations into a series of possible executions where all prior inputs and generated outputs have been fully processed. The set of operator states at the end of a stage reflectsg the whole execution history, therefore, it can be solely used for a snapshot. The core idea behind our algorithm is to create identical snapshots with staged snapshotting while keeping a continuous data ingestion.

In our approach, stages are emulated in a continuous dataflow execution by special barrier markers injected in the input data streams periodically that are pushed throughout the whole execution graph down to the sinks. Global snapshots are incrementally constructed as each task receives the barriers indicating execution stages. We further make the following assumptions for our algorithm: 
\begin{itemize}
  \item Network channels are quasi-reliable, respect a FIFO delivery order and can be \emph{blocked} and \emph{unblocked}. When a channel is \emph{blocked} all messages are buffered but not delivered until it gets \emph{unblocked}.
  \item Tasks can trigger operations on their channel components such as \emph{block}, \emph{unblock} and \emph{send} messages. \emph{Broadcasting} messages is also supported on all output channels.
  \item Messages injected in source tasks (i.e. stage barriers) are resolved into a ``Nil'' input channel.
\end{itemize}

\noindent \textbf{The ABS Algorithm \ref{carbonefora}  proceeds as follows (depicted in Fig. \ref{abs-dag}):} A central coordinator periodically injects stage barriers to all the sources. When a source receives a barrier it takes a snapshot of its current state, then broadcasts the barrier to all its outputs (Fig.\ref{abs-dag}(a)). When a non-source task receives a barrier from one of its inputs, it blocks that input until it receives a barrier from all inputs (line \ref{lst:line:block}, Fig.\ref{abs-dag}(b)). When barriers have been received from all the inputs, the task takes a snapshot of its current state and broadcasts the barrier to its outputs (lines \ref{lst:line:barrier}-\ref{lst:line:snap}, Fig.\ref{abs-dag}(c)).  Then, the task unblocks its input channels to continue its computation (line \ref{lst:line:unblock}, Fig.\ref{abs-dag}(d)). The complete global snapshot $G^*=(T^*,E^*)$ will  consist solely of all operator states $T^*$ where $E^* = \emptyset$.

\noindent \textbf{Proof Sketch \footnote{We omit the formal proof of the algorithms in this version of the paper due to space limitations}: }As mentioned earlier a snapshot algorithm should guarantee \emph{termination} and \emph{feasibility}. 

\noindent \emph{Termination} is guaranteed by the channel and acyclic execution graph properties. Reliability of the channels ensures that every barrier sent will eventually be received as long as the tasks are alive. Furthermore, as there is always a path from a source, every task in a DAG topology will eventually receive barriers from all its input channels and take a snapshot.

\noindent For \emph{feasibility} it suffices to show that the operator states in a global snapshot reflect only the history of records processed up to the last stage. This is guaranteed by the FIFO ordering property of the channels and the blocking of input channels upon barriers ensuring that no post-shot records of a stage (records succeeding a barrier) are processed before a snapshot is taken.

\begin{figure*}[t]
\centering
\includegraphics[width=7in]{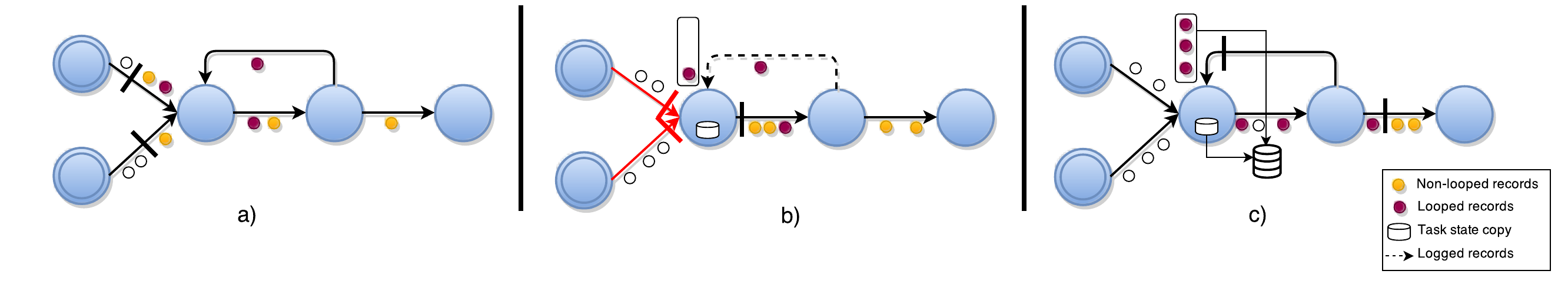}
\vspace{-1em}
\caption{Asynchronous barrier snapshots for cyclic graphs}
\vspace{-0.5em}
\label{abs-cyclic}
\end{figure*}
\label{proof1}

\begin{algorithm}[h]
\caption{Asynchronous Barrier Snapshotting for Cyclic Execution Graphs} 
\label{carbonefora2}
\begin{algorithmic}[1]
\Upon[Init]{input\_channels, backedge\_channels, output\_channels, fun, init\_state}
  \State $state :=  init\_state;$ $marked := \emptyset;$
  \State $inputs := input\_channels;$ $logging := False;$
  \State $outputs := output\_channels;$ \emph{udf} $:= fun;$
  \State $loop\_inputs :=$ $backedge\_channels$;
  \State $state\_copy := Nil;$ $backup\_log := [];$

\EndUpon
\Upon[receive]{input, $\langle barrier \rangle$}
  \State $marked := marked \cup \{input\};$
  \State $regular := inputs \setminus loop\_inputs;$
  \If{$input \neq Nil$ AND $input \notin loop\_inputs$}
    \State \textbf{trigger} $\langle$block $\mid$ input$\rangle;$
  \EndIf
  \If{$\lnot logging$ AND $marked = regular$}
    \State $state\_copy := state; logging := True;$\label{lst:line:statecopy}
\State \textbf{broadcast} $\langle$send $\mid$ outputs, $\langle barrier \rangle$$\rangle$;
     \ForEach[inputs]{input}
      \State \textbf{trigger} $\langle$unblock $\mid$ input $\rangle;$
    \EndForEach      
      \EndIf
  \If{$marked = input\_channels$}
   \State \textbf{trigger} $\langle$snapshot $\mid$ \{state\_copy, backup\_log\}$\rangle;$
   \State $marked := \emptyset;logging := False;$
   \State $state\_copy := Nil;backup\_log := [];$
  \EndIf
\EndUpon
\Upon[receive]{input, msg}
  \If{$logging$ AND $node \in loop\_inputs$}
    \State $backup\_log := backup\_log::[input]$; \label{lst:line:backuplog}
  \EndIf
    \State $\{state',out\_records\} :=  $ \emph{udf}$(msg, state);$
    \State $state := state';$
    \ForEach[out\_records]{\{output, out\_record\}}
      \State \textbf{trigger} $\langle$send $\mid$ output, out\_record$\rangle;$
    \EndForEach
\EndUpon
\end{algorithmic}
\end{algorithm}

\subsection{ABS for Cyclic Dataflows}

In the presence of directed cycles in the execution graph the ABS algorithm presented before would not terminate resulting into a deadlock, as tasks in a cycle would wait indefinitely to receive barriers from all their inputs. Additionally, records that are arbitrarily in transit within cycles would not be included in the snapshot, thus violating feasibility. It is therefore needed to consistently include all records generated within a cycle in the snapshot for feasibility and to put these records back in transit upon recovery. Our approach to deal with cyclic graphs extends the basic algorithm without inducing any additional channel blocking as seen in Algorithm \ref{carbonefora2}. First, we identify \emph{back-edges} $L$ on loops in the execution graph by static analysis. From control flow graph theory a \emph{back-edge} in a directed graph is an edge that points to a vertex that has already been visited during a depth-first search. The execution graph $G(T,E \setminus L)$ is a DAG containing all tasks in the topology. From the perspective of this DAG the algorithm operates as before, nevertheless, we additionally apply downstream backup of records received from identified \emph{back-edges} over the duration of a snapshot. This is achieved by each task $t$ that is a consumer of back-edges $L_t \subseteq I_t,L_t$ creating a backup log of all records received from $L_t$ from the moment it forwards barriers until receiving them back from $L_t$. Barriers push all records in transit within loops into the downstream logs, so they are included once in the consistent snapshot.

\noindent \textbf{The ABS Algorithm \ref{carbonefora2} proceeds as follows (depicted in Fig. \ref{abs-cyclic}) in more detail:}  Tasks with back-edge inputs create a local copy of their state once all their regular ($e \notin L$) channels have delivered barriers (line \ref{lst:line:statecopy}, Fig.\ref{abs-cyclic} (b)). Moreover,  from this point they log all records delivered from their back-edges until they receive stage barriers from them (line \ref{lst:line:backuplog}). This allows, as seen in  Fig. \ref{abs-cyclic}(c), for all pre-shot records that are in transit within loops to be included in the current snapshot.  Mind that the final global snapshot $G^*=(T^*, L^*)$ contains all task states $T^*$ and only the back-edge records in transit $L^* \subset E^*$.

\noindent \textbf{Proof Sketch: } Again, we need to prove that \emph{termination} and \emph{feasibility} are guaranteed in this version of the algorithm. 

\noindent  As in \ref{proof1} \emph{termination} is guaranteed, because every task will eventually receive barriers from all its inputs (including the back-end channels) and complete its snapshot. By broadcasting the barrier as soon as receiving it from all regular inputs, we avoid the deadlock condition mentioned previously.

\noindent The FIFO ordering property still holds for back-edges and the following properties prove \emph{feasibility} . (1) Each task state included in the snapshot is a state copy of the respective task taken before processing any post-shot events from barriers received on regular inputs. (2) The downstream log included in the snapshot is complete and contains all pending post-shot records prior to barriers received on back-edges due to FIFO guarantees.

\section{Failure Recovery}
\label{recov}
While not being the main focus of this work, a working failure recovery scheme motivates the application of our snapshotting approach. Thus, we provide a brief explanation here regarding its operation. There are several failure recovery schemes that work with consistent snapshots. In its simplest form the whole execution graph can be restarted from the last global snapshot as such:  every task $t$  (1) retrieves from persistent storage its associated state for the snapshot $s_t$ and sets it as its initial state, (2) recovers its backup log and processes all contained records, (3) starts ingesting records from its input channels.

\begin{figure}[t!]
\centering
\includegraphics[width=2.2in]{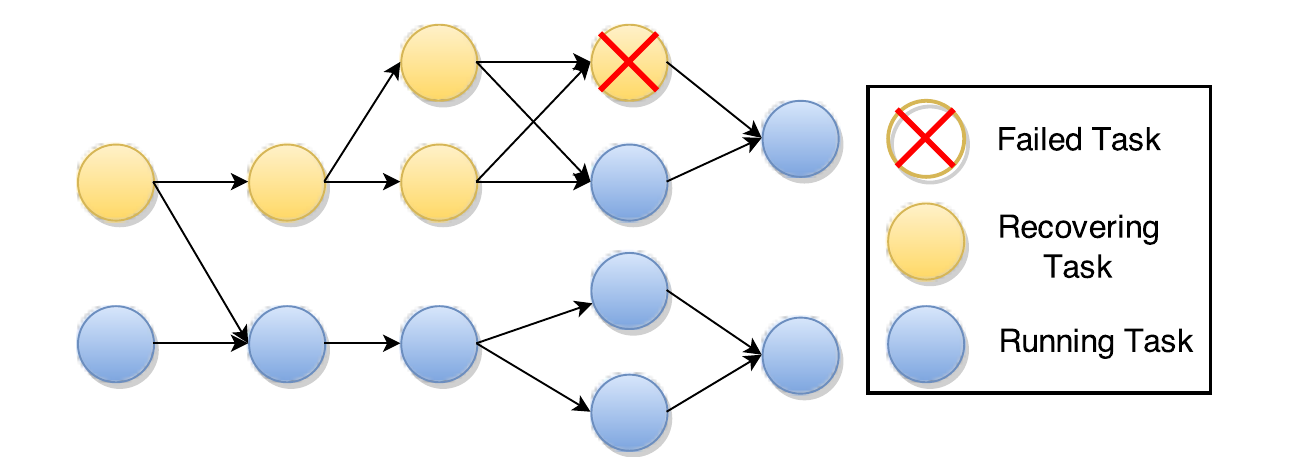}
\vspace{-0.3em}
\caption{Upstream Dependency Recovery Scheme}
\label{recovery}
\end{figure}

A partial graph recovery scheme is also possible, similarly to TimeStream \cite{qian2013timestream}, by  rescheduling only upstream task dependencies (tasks that hold output channels to the failed tasks) and their respective upstream tasks up to the sources. An example recovery plan is shown in Fig. \ref{recovery}. In order to offer \emph{exactly-once} semantics, duplicate records should be ignored in all downstream nodes to avoid recomputation. To achieve this we can follow a similar scheme to SDGs \cite{fernandez2014making} and mark records with sequence numbers from the sources, thus, every downstream node can discard records with sequence numbers less than what they have processed already. 

\begin{figure*}[t!]
    \centering
    \begin{minipage}[t]{2in}
        \centering
        \includegraphics[width=2in,height=1.5in]{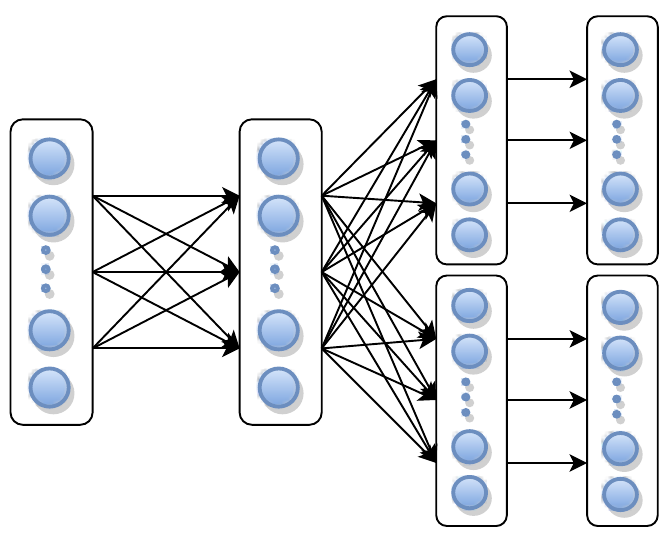}
        \caption{Execution topology used for evaluation}
        \label{fig:exp-topology}
    \end{minipage}
    ~ 
    \begin{minipage}[t]{2.3in}
        \centering
        \includegraphics[width=2.3in]{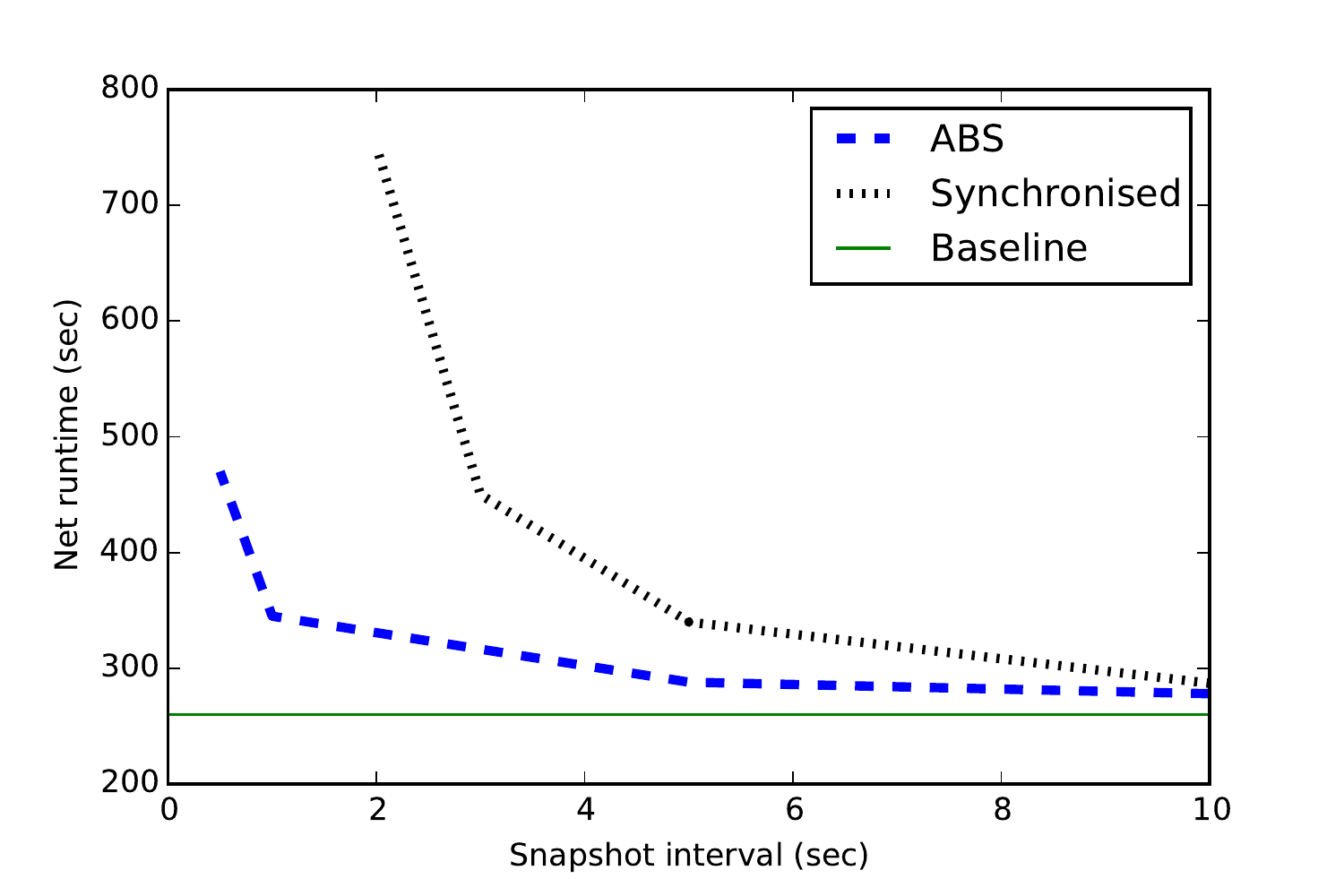}
        \caption{Runtime impact comparison on varying snapshot intervals}
        \label{fig:exp-perf}
    \end{minipage}
    ~
    \begin{minipage}[t]{2.3in}
        \centering
        \includegraphics[width=2.3in]{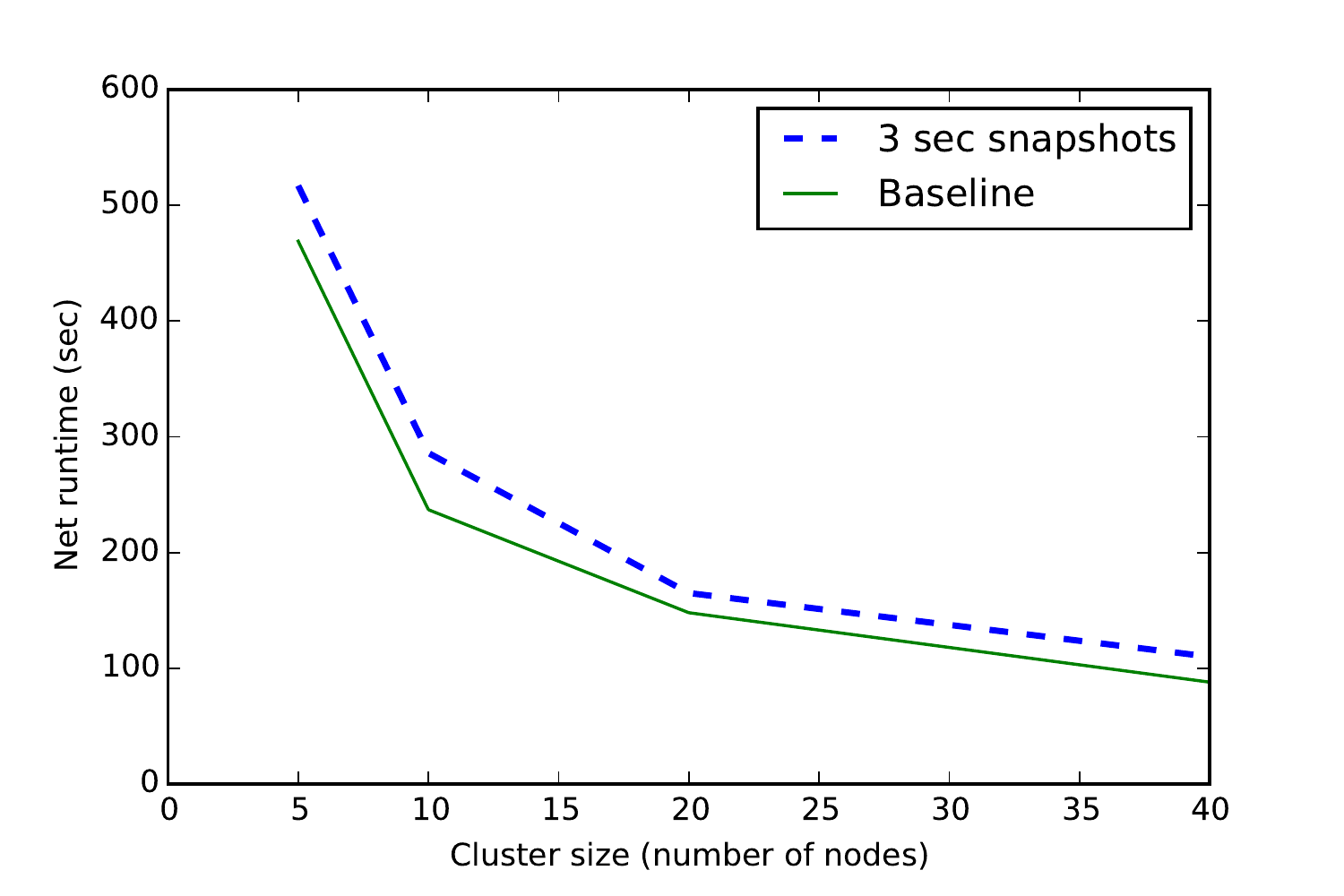}
        \caption{Runtime with ABS on different cluster sizes}
        \label{fig:exp-scale}
    \end{minipage}
\end{figure*}

\section{Implementation}
\label{impl}
We contributed the implementation of the ABS algorithm to Apache Flink in order to provide exactly-once processing semantics for the streaming runtime. In our current implementation blocked channels store all incoming records on disk instead of keeping them in memory to increase scalability. While this technique ensures robustness, it increases the runtime impact of the ABS algorithm.

In order to distinguish operator state from data we introduced an explicit \emph{OperatorState} interface which contains methods for updating and checkpointing the state. We provided \emph{OperatorState} implementations for the stateful runtime operators supported by Apache Flink such as offset based sources or aggregations.

Snapshot coordination is implemented as an actor process on the job manager that keeps a global state for an execution graph of a single job. The coordinator periodically injects stage barriers to all sources of the execution graph. Upon reconfiguration, the last globally snapshotted state is restored in the operators from a distributed in-memory persistent storage. 

\section{Evaluation}
\label{eval}

The goal of our evaluation is to compare the runtime overhead of ABS with the globally synchronised snapshot algorithm employed in Naiad \cite{murray2013naiad} and also to test the scalability of the algorithm to larger number of nodes.

\subsection{Experiment Setup}

The execution topology (Fig. \ref{fig:exp-topology}) used for evaluation consists of 6 distinct operators with parallelism equal to the number of cluster nodes, which translates into a 6*\emph{cluster size} task vertices. The execution contains 3 full network shuffles in order to accentuate the possible impact of channel blocking in ABS. Sources generate a total of 1 billion records which are distributed uniformly among the source instances. The state of the operators in the topology were per-key aggregates and source offsets. The experiments were run on Amazon EC2 cluster using up to 40 m3.medium instances. 

We measured the runtime overhead of our evaluation job running under different snapshotting schemes, namely ABS and synchronised snapshotting \cite{murray2013naiad} with varying snapshot intervals. We implemented the synchronous snapshotting algorithm used in Naiad \cite{murray2013naiad} on Apache Flink in order to have identical execution backend for the comparison. This experiment was run using a 10 node cluster. To evaluate the scalability of our algorithm, we processed a fixed amount of input records (1 billion) while increasing the parallelism of our topology from 5 up to 40 nodes.

\subsection{Results}

In Fig. \ref{fig:exp-perf} we depict the runtime impact of the two algorithms against the baseline (no fault tolerance). The large performance impact of synchronous snapshotting is especially visible when the interval of snapshotting is small. That is due to the fact that the system spends more time not processing any data, in order to obtain global snapshots. ABS has a much lower impact on the runtime as it runs continuously without blocking the overall execution, while maintaining a rather stable throughput rate. For larger snapshot intervals the impact of the synchronised algorithm is less significant since it operates in bursts (of 1-2 sec in our experiments) while letting the system run at its normal throughput during the rest of its execution. Nevertheless, bursts can often violate SLAs in terms of real-time guarantees for many applications that are latency-critical, such as intrusion detection pipelines. Thus, such applications would further benefit by the performance of ABS. In Fig. \ref{fig:exp-scale} we compare the scalability of the topology running ABS with 3 second snapshot intervals against the baseline (without fault tolerance). It is clear that both the baseline job and ABS achieved linear scalability.

\section{Future Work and Conclusion}
\label{futurework}

In future work we are planning to explore the possibilities of further lowering the impact of ABS by decoupling snapshotting state and operational state. That allows for purely asynchronous state management as tasks can continuously process records while persisting snapshots. In such scheme there is also the need to synchronise pre-shot and post-shot records with the respective state which can be resolved by marking the records depending on the snapshot they belong to. As this approach would increase the computational, space and network I/O requirements of the algorithm, we plan to compare its performance to our current ABS implementation. 
Finally, we plan to investigate different recovery techniques that maintain exactly-once semantics while minimising the need for reconfiguration by operating on a per-task granularity. 

In summary,  we focused on the problem of performing periodic global snapshots on distributed dataflow systems. We introduced ABS, a new snapshotting technique that achieves good throughput. ABS is the first algorithm that considers the minimum state possible for acyclic execution topologies. Furthermore, we extend ABS to work with cyclic execution graphs by storing only the records that need to be reprocessed upon recovery. We implemented ABS on Apache Flink and evaluated our approach against synchronous snapshotting. At this early stage ABS shows good results, having low impact on the overall execution throughput with linear scalability. 
\bibliographystyle{abbrv}
\bibliography{references}

\balance
\end{document}